\documentclass[12pt]{article}
\usepackage[top=3cm,bottom=3cm,left=3cm,right=3cm]{geometry}

\usepackage{multirow,tabularx}
\usepackage{amsmath,amsthm,amssymb,graphicx}

\usepackage[colorlinks=true,urlcolor=blue, linkcolor =blue,citecolor=blue]{hyperref}
\usepackage[width=.9\textwidth, labelfont=bf,
            textfont=it]{caption}

\usepackage{algorithm,algorithmicx,algpseudocode}
\algblockdefx{MRepeat}{EndRepeat}{\textbf{Repeat}}{}
\algnotext{EndRepeat}

\makeatletter
\renewcommand{\@biblabel}[1]{}
\renewenvironment{thebibliography}[1]
     {\section*{\refname}%
      \@mkboth{\MakeUppercase\refname}{\MakeUppercase\refname}%
      \list{}%
           {\labelwidth=0pt
            \labelsep=0pt
            \leftmargin1em
            \itemindent=-1em
            \advance\leftmargin\labelsep
            \@openbib@code
            }%
      \sloppy
      \clubpenalty4000
      \@clubpenalty \clubpenalty
      \widowpenalty1000%
      \sfcode`\.\@m}
\makeatother
\usepackage{breakcites}

\numberwithin{equation}{section}

\theoremstyle{definition}

\theoremstyle{remark}

\allowdisplaybreaks

\begin{document}

\title{\bf Boosting heritability: estimating the genetic component of phenotypic variation with multiple sample splitting}
\author{The Tien Mai$^{(1)}_{\footnote{Corresponding author: t.t.mai@medisin.uio.no}}$, Paul Turner$^{(3),(4)}$ and Jukka Corander$^{(1),(2)}$}

\date{
\small
$^{(1)}$ Oslo Centre for Biostatistics and Epidemiology, Department of Biostatistics, 
\\
University of Oslo, Norway.
\\
$^{(2)}$ Department of Mathematics and Statistics, University of Helsinki, Finland.
\\
$^{(3)}$ Cambodia-Oxford Medical Research Unit, Angkor Hospital for Children, 
\\
Siem Reap, Cambodia.
\\
$^{(4)}$Centre for Tropical Medicine and Global Health, Nuffield Department of Medicine,
\\
University of Oxford, UK.
}
\maketitle

\begin{abstract}
\textit{ } \vspace*{-1cm}
\\
{\bf Background}:
Heritability is a central measure in genetics quantifying how much of the variability observed in a trait is attributable to genetic differences. Existing methods for estimating heritability are most often based on random-effect models, typically for computational reasons. The alternative of using a fixed-effect model has received much more limited attention in the literature. 
\vspace*{.2cm}
\\
{\bf Results}:
In this paper, we propose a generic strategy for heritability inference, termed as \textit{``boosting heritability"}, by combining the advantageous features of different recent methods to produce an estimate of the heritability with a high-dimensional linear model. Boosting heritability uses in particular a multiple sample splitting strategy which leads in general to a stable and and accurate estimate. We use both simulated data and real antibiotic resistance data from a major human pathogen, \textit{Sptreptococcus pneumoniae}, to demonstrate the attractive features of our inference strategy.
\vspace*{.2cm}
\\
{\bf Conclusions}: Boosting is shown to offer a reliable and practically useful tool for inference about heritability. 
\end{abstract}


\section{Introduction}
Whereas genome-wide association studies (GWAS) represent the primary tool for determining the genetic basis of a phenotype/trait of interest, quantifying the contribution of genetic factors to the variation of a phenotype plays in addition an important role in many studies. For this purpose, \textit{heritability} is a crucial quantity \cite{falconer1960introduction,lynch1998genetics} and it is defined (in the narrow-sense) as the proportion of the variance of a phenotype explained by the (additive) genetic factors.

Current studies of heritability in the literature have usually been carried out in the linear mixed-effect model framework \cite{bulik2015ld,yang2010common}. In this framework, the effect sizes of genetic markers, usually SNPs, are assumed to be independent and identical distributed random variables, and often the normal distribution (with 0-mean) is used for computational reasons. The maximum likelihood and method of moments are the most widely used methods for heritability inference for this family of models \cite{yang2010common,golan2014measuring,bulik2015ld,
zhou2017unified,bonnet2016heritability,speed2017reevaluation}. 

Some comparisons of different methods for estimating heritability have been recently conducted, for example, in \cite{zhou2017unified,evans2018comparison,
weissbrod2018estimating,gorfine2017heritability}. However, these works compare the performance of different methods on different datasets without paying much attention to the actual model specification. Since heritability is a concept detailing the additive variance of a trait which is in a certain sense based on a statistical model, heritability estimation is consequently dependent on the specified model \cite{zaitlen2012heritability}. For example, as reported in \cite{evans2018comparison}, there is a sizeable difference in the estimated heritability of schizophrenia $\hat{h}^2_{SNP} $ that equals 0.56 according to \cite{bulik2015ld} and only 0.23 according to \cite{lee2013genetic}. These estimates have a very different interpretation also qualitatively and they disagree most likely because they are based on different statistical models of heritability.

In this paper, we focus on the high-dimensional linear regression model with fixed effects, where no distributional assumption on the effect sizes is made. Although limited from the computational perspective due to the extremely high-dimensional data in GWAS, high-dimensional linear regression is a natural model for GWAS in modelling the whole-genome level contributions of genetic variation. The benefit of this model over the classical univariate approach in GWAS has been demonstrated for example in \cite{wu2009genome,brzyski2017controlling}. The study of heritability estimation with fixed-effect models has been started relatively recently and it has not yet gained a wide-spread attention. A method of moments approach is proposed in \cite{dicker2014variance}, a convex optimization strategy is suggested in \cite{janson2017eigenprism} through a singular value decomposition, maximum likelihood estimation is studied in \cite{dicker2016maximum}, and some adaptive procedures have also been theoretically studied in \cite{verzelen2018adaptive}. However, to our knowledge, a systematic numerical comparison of these different methods for estimating heritability has not been made yet. 

Some two-step procedures based on high-dimensional regularized regression have been introduced in \cite{gorfine2017heritability,li2019reliable} that provide an insight to obtain more reliable and stable estimates of heritability. In brevity, this approach is based on splitting the data into two subsets. In the first step, variable selection is employed through a sparsity inducing regularization on one subset to select the relevant covariates. In the second step, these selected covariates are used to estimate heritability from the other subset of data. The selection step is useful to consider only a subset of the covariates that contribute to the variability of the trait (the response). Moreover, splitting the sample is done to avoid doing variable selection and heritability estimation on the same data which can cause overestimate \cite{li2019reliable}. Although promising, this approach depends crucially on the particular partition used to split the data, which can lead to unstable estimates.

To achieve more reliable results, we propose to use a multiple sample splitting procedure so that different structures in the sample are presented in both selection and estimation steps with a sufficiently high probability \cite{meinshausen2009p,fan2012variance}. Based on this idea, we present a general framework called ``boosting heritability" which allows a user to plug-in their own favourite method of variable selection and/or heritability estimation. By repeating sample splitting, one can also obtain various estimates of the heritability and thus provide a meaningful interval of the estimated values.

To demonstrate our framework, we apply the procedure to bacterial GWAS for estimating the heritability of antibiotic resistant phenotypes. While there are numerous works concerning estimating heritability in human GWAS, the topic has not yet been considered widely in bacteria, for the only prominent example see \cite{lees2017genome}. This is partly because bacterial GWAS poses unique challenges compare to studies with human or animal DNA, stemming from more limited recombination and highly structured populations that result in substantial linkage disequilibrium across whole chromosomes. 

The paper is structured as follows. In Section~\ref{sec:model} we present the linear model that relates a trait with a genotype matrix, then narrow-sense heritability is defined together with some discussion regarding the fixed-effect vs. random-effect approach for estimation. In Section~\ref{sec.boosther},  we introduce our ``boosting heritability" procedure. Results from a simulation study comparing the different methods as components of the framework presented in Section~\ref{sec.simulation} and the application to antibiotic resistance phenotypes are presented in Section~\ref{sc_realdata}. We conclude and discuss our results in the final section.

\section{Model and definition}
\label{sec:model}
\paragraph{Notations:} Here, we introduce the main notations used in the paper. The $\ell_q $ norm $ (0<q<+\infty)$ of a vector $x \in \mathbb{R}^d$ is defined by $ \|x \|_q = (\sum_{i=1}^d |x_i|^q )^{1/q} $. For a matrix $A\in \mathbb{R}^{n\times m} $, $A_{i\cdot} $ denotes its $i$-th row and $A_{\cdot j} $ denotes its $j$-th column. For any index set $S \subseteq \{1,\ldots,d\} $, $x_S$ denotes the subvector of $x$ containing only the components indexed by $S$, and $A_S$ denotes the submatrix of $A$ forming by columns of $A$ indexed by $S$.

\subsection{Model}
Given a phenotype/trait $ y $ of $n$ samples that is modelled as a linear combination of $ p $ genetic covariates $ X_{\cdot j} $ and an error term (environmental and unmeasured genetic effects)
\begin{align}
\label{linearmodel}
y_{i} = X_{i\cdot} \beta + \varepsilon_{i}, i = 1,\ldots,n
\end{align}
where $ X_{i\cdot} $ are independent and identically distributed (i.i.d)  with distribution $\mathcal{N}(0, \Sigma) $ and are independent of $ \varepsilon_{i} \sim \mathcal{N} (0, \sigma^2_{\varepsilon}) $. 

Here we focus on the fixed effects encoded by $\beta$ and assume that the genetic covariates $X$ are random variables. Conversely, in the majority of works in the heritability literature assume that elements of $\beta$ are considered as i.i.d. random variables following a Gaussian distribution i.e $\beta_j \overset{i.i.d}{\sim} \mathcal{N} (0, \sigma^2_{\beta}) $, while the genetic covariates $X$ are assumed fixed.

\subsection{Heritability}
Under the model \eqref{linearmodel}, we have for the $ i$-th observation that
\begin{align*}
{\rm Var}(y_i) = {\rm Var}(X_{i\cdot}\beta ) + \sigma_{\varepsilon}^2 = \beta^\top \Sigma \beta + \sigma_{\varepsilon}^2 .
\end{align*}
We are interested in estimating (the narrow-sense) heritability for $y$ defined as
\begin{align}
\label{heritability:formula}
h^2 = \frac{ \beta^\top \Sigma \beta }{\beta^\top \Sigma \beta+ \sigma_{\varepsilon}^2 } .
\end{align}

Technically, heritability is a quantitative measure that expresses how much of the population variability present in a trait is due to genetic differences. Moreover, estimating heritability can assist in modelling the underlying genetic architecture. A heritability close to zero implies that environmental factors cause most of the variability of the trait. In contrast, a heritability close to 1 indicates that the variability of the trait is nearly exclusively caused by the differences in genetic factors. 

As we have the relation
$$
\mathbb{E} [\| y\|_2^2/n ]   
= 
{\rm Var} (y)
=
\beta^\top \Sigma \beta + \sigma_{\varepsilon}^2,
$$
one can use $\| y\|_2^2/n $ as an unbiased estimator for the denominator of the heritability. Further, one can re-write \eqref{heritability:formula} as 
\begin{align}
\label{herit:formula-re-write}
h^2 = 1 -  \frac{\sigma_{\varepsilon}^2 }{{\rm Var}(y)  } 
\end{align}
and use an estimate of the noise-variance $\hat{\sigma}_{\varepsilon}^2 $ (see e.g \cite{reid2016study}) to estimate $h^2 $ rather than directly estimate the genetic variance $\beta^\top \Sigma \beta $ (which requires an estimate of the covariance matrix and the effect sizes). 
\\
However, it is worth noting that as a bi-product from GWAS analysis when using a multivariate regression approach, such as the Elastic net discussed below, one would already have the estimated effect sizes corresponding to the selected covariates. Using these effect sizes to estimate the heritability would bring insight on the heritability corresponding to the selected covariates and thus clearly provide useful ways to understand the genetic architecture of a trait.

\subsection*{Contrasting the fixed and random effects}
In GWAS the true number of causal loci reported tend to be comparatively small compared with the number of putative genetic markers $p $, which is usually in the order of hundreds of thousands at minimum. Assume that the true effect size $\beta$ has $s\ll p $ non-zero entries. In the random-effect model, a further assumption is made concerning these non-zero entries such that they are i.i.d Gaussian $\mathcal{N} (0, \sigma^2_\beta)$. Under this random effect assumption, the heritability is defined \cite{bonnet2015heritability,li2019reliable} as  
$
s \sigma^2_\beta / (s \sigma^2_\beta + \sigma_{\varepsilon}^2) .
$

However, when employing the random-effect assumption, most methods do not use the sparsity constraint. This leads to the target heritability being estimating is
$
p \sigma^2_\beta / (p \sigma^2_\beta + \sigma_{\varepsilon}^2)
$
and the resulting estimate of heritability may thus be inaccurate. Moreover, the LD structure, an important concept that represents the correlation structure of the covariates, is not directly addressed in the formula of heritability in random-effect model, which can make the estimate unjustifiable, e.g see \cite{speed2017reevaluation,speed2019sumher}. Several attempts have been done recently to take into account the sparsity constraint within the random-effect model and some promising results have been reported in \cite{bonnet2015heritability,bonnet2018improving,li2019reliable}.

\section{Boosting heritability estimation}
\label{sec.boosther}

\subsection{Related works and motivation}
As the number of biomarkers can be very large, it is natural to first apply some variable selection or variable screening methods to remove the irrelevant variables from the actual heritability estimation phase. This kind of a post-selection approach has been proposed in the literature, more specifically for the fixed-effect model \cite{gorfine2017heritability,li2019reliable}.

The HERRA method proposed in \cite{gorfine2017heritability} is based on a screening method (e.g. as in \cite{fan2008sure}) to reduce the number of covariates below the sample size. Given the remaining covariates, the sample is randomly divided  into two equally sized parts. A lasso-type estimator is employed on the first subset to select a small number of important variables. After that, the least squares estimator is used on the second subset of data using only the selected covariates (from the lasso-type estimator) to get an estimate of the noise-variance. The role of the first and second subsets are switched to obtain another estimate of the noise-variance. Finally, heritability is calculated as in the formula \eqref{herit:formula-re-write} where the noise-variance is the mean of the two estimated noise-variances.

Another ``two-stage" approach with sample-splitting has also been proposed in the paper \cite{li2019reliable}. The data is randomly split into two disjoint equal sample size. On one half of the data, they use a sparse regularization method based on Elastic net to first select the relevant variables. Then, on the other half of the data, they only use the selected variables to estimate the heritability through a method of moments based approach \cite{dicker2014variance}. 

Both these approaches clearly suffer from some limitations. Firstly, when the number of covariates is very large, it is expensive to fit a sparse regularization directly as in the ``two-stage" approach described above. Using a screening method, as in HERRA, to reduce the dimension of the problem is thus a pragmatic approach for applications. However, as the true number of causal biomarkers is not known, as well as their LD structure is not given, reducing the number of variables below the sample size (as in HERRA) introduces another problem from the practical perspective. Secondly, it is clear that both of these approaches crucially depend on the particular sample splitting employed. One can avoid this dependence  by performing the sample splitting and inference procedure many times (e.g. 100 times) and aggregating the corresponding results  \cite{meinshausen2009p,fan2012variance}. This is to ensure that the different latent structures possibly residing in the sample are properly taken into account in both the selection and estimation steps. 

The idea of aggregating different estimates to yield an estimate with improved statistical properties is the central feature of the generic boosting approach widely used in machine learning, such as AdaBoost \cite{freund1996experiments}. The multiple sample splitting approach has previously been proposed in statistics community as in \cite{meinshausen2009p,fan2012variance}, and successfully used in GWAS \cite{renaux2020hierarchical,buzdugan2016assessing}.

\subsection{Boosting heritability:  multi sample splitting and  aggregation of heritability}

We propose a strategy that uses multiple sample splitting to estimate heritability, called Boosting heritability detailed in Algorithm~\ref{Boostingheritability}.

\begin{algorithm}{}
\caption{Boosting heritability}
\begin{algorithmic}[1]
\State \textbf {Step 0:}  Using a screening method, such as a marginal-type sure independent screening \cite{fan2008sure}, to remove 25\% irrelevant covariates. This step aims at reduce the ultra-high dimension to a more manageable level. 

\MRepeat $\, B$ times from step 1 to step 4,
 \State  \textbf{Step 1:} With the remaining covariates, divide the sample uniformly at random into two equal parts. 
 \State \textbf{Step 2:} On the first part of the data, use Elastic net to select the important covariates.
 \State \textbf{Step 3:} Then, on the second subset with only selected covariates from Step 2, estimate the heritability by using a method presented in Section \ref{sec.herit.method}.
 \State \textbf{Step 4:} Repeat Step 2 and Step 3 by changing the role of the first and second subset.
\EndRepeat
\State \textbf{Final} $\rightarrow $ The final heritability estimate is the mean of the estimated heritabilities at each repeat.          
\end{algorithmic}
\label{Boostingheritability}
\end{algorithm}

It is noted that the initial step ({\bf Step 0}) is a screening step that can use a simple measure of association, such as the sample correlation, to remove covariates that are only weakly correlated with the trait of interest. This step is similar to the one used in HERRA~\cite{gorfine2017heritability} and in \cite{bonnet2018improving}, however, we do not propose to reduce the number of covariates below the actual sample size. This is motivated by the fact for real data we do not know the true number of causal variates as well as the correlation structure of the variables. If too many covariates are removed, this can have a detrimental effect on the subsequent steps in the estimation procedure. Moreover, the initial screening step can be seen as optional, and necessary only for situations where the high dimensionality of the covariate space makes regularized model fitting tedious or practically impossible for practical purposes.  

\begin{figure}[!ht]
\centering
\includegraphics[height=8cm, width=13cm]{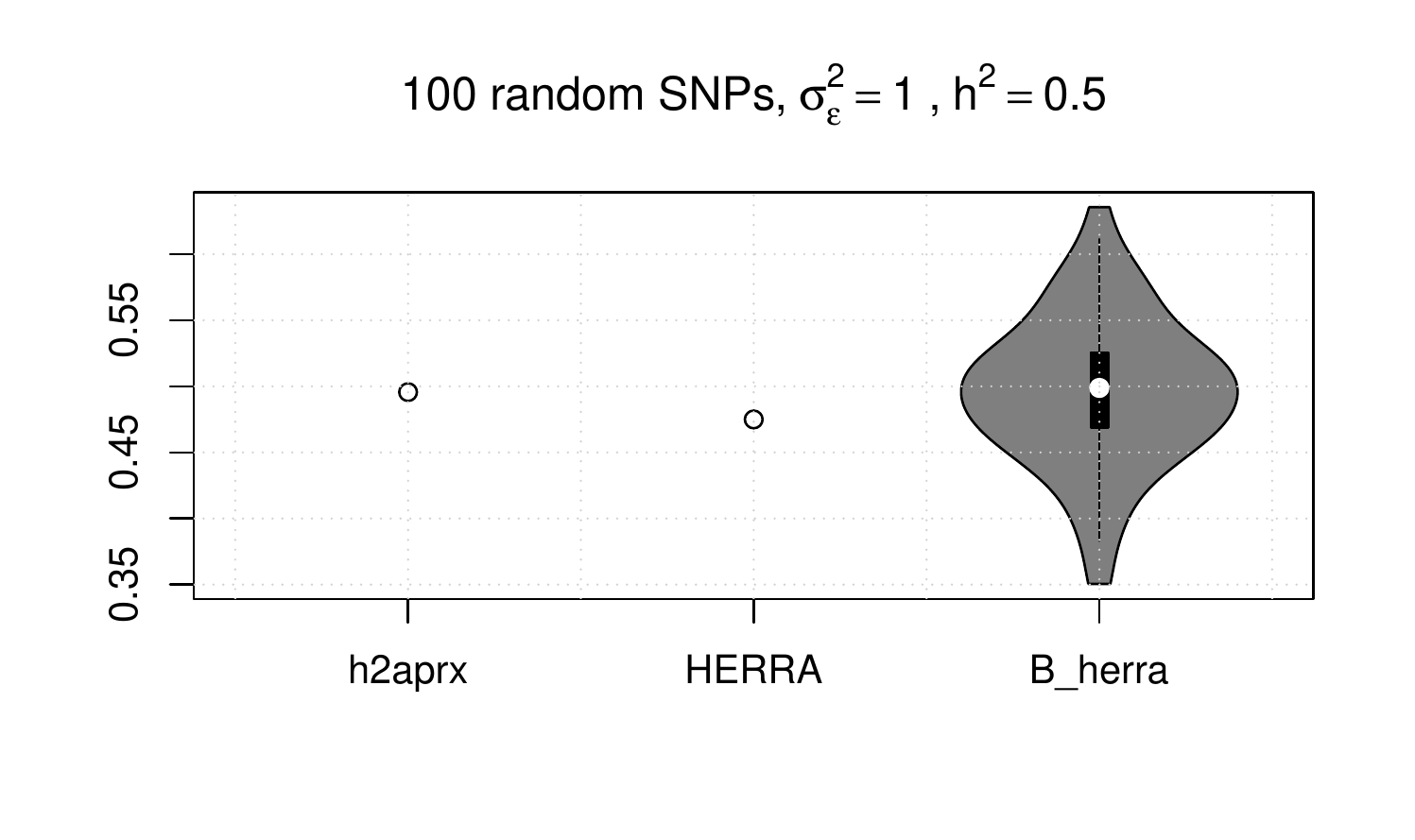}
\vspace*{-1.5cm}
\caption{A violin plot for estimates of heritability from the simulation with MA data with 100 random covariates chosen as causal. We obtain an interval of heritabilities through the multiple sample splitting method (B\_herra, with $B=50$). See Section \ref{sec.simulation}.}
\label{multiple_herit_estimates}
\end{figure}

The sample splitting performed in {\bf Step 1} is a useful method that can help to avoid overfitting when variable selection and subsequent estimation is considered \cite{fan2012variance,
buzdugan2016assessing,li2019reliable}. {\bf Step 2} corresponds to a variable selection phase where we suggest to use Elastic Net as a default alternative, given its ability to deal with highly correlated covariates. Switching the roles of the data subsets help us to obtain a more stable estimate of the heritability. Note that by repeating sample splitting, various estimates of the heritability are obtained and thus provide a meaningful interval of the estimated values (for example see Figure \ref{multiple_herit_estimates}).

We note that the main cost for Boosting Heritability procedure is fitting a penalized regression ({\bf Step 2}) for variable selection in the setting where $ p \gg n$. However, fast computation methods for penalized regression on large GWAS data have been recently proposed see e.g \cite{qian2020fast}. Moreover, the $B$ repetitions can be easily implemented in parallel. When the trait of interest is dichotomous, one can use the Robertson transformation \cite{dempster1950heritability} to transfer the heritabilty calculated on an observed scale (on 0 or 1) to a heritability on the liability scale. As we largely follow the approach presented in the HERRA method, the details for obtaining heritability for a binary trait can be found in \cite{gorfine2017heritability} or in \cite{lee2011estimating}.

\subsubsection{Plug-in Lasso type estimators for heritability}
\label{sec.herit.method}
From the formula of heritability  \eqref{heritability:formula}, direct approaches to estimate heritability can be obtained using estimates of the effect sizes $\beta$ and of the covariance matrix. By using a lasso type method, one can obtain the non-zero estimated effect sizes of the selected covariates, and one can also use these covariates to obtain an sample covariance matrix. More precisely, let $S = \left\lbrace j : \hat{\beta} \neq 0 \right\rbrace $ where $ \hat{\beta} $ is an estimate from a lasso-type method, we can calculate the heritability as in equation \eqref{heritability:formula} with $ \hat{\Sigma}_S = X_S X_S^\top/(n-1) $,
\begin{align*}
\hat{h}^2 = \frac{ \hat{\beta}_S^\top \hat{\Sigma}_S \hat{\beta}_S  }{{\rm Var}(y)} .
\end{align*}

The elastic net has been shown to be especially useful when the variables are dependent \cite{zou2005regularization} (LD structure), which is often the case with genetic marker data and this feature is especially highlighted in bacterial genome data. The corresponding estimator is defined as
\begin{align*}
\hat{\beta}_{Enet} := \arg \min_{\beta} \frac{1}{n} \sum_{i=1}^{n}  \ell(y_i,\beta^T x_i) + \lambda\left[0.5(1-\alpha)||\beta||_2^2 + \alpha ||\beta||_1\right] .
\end{align*}
Here \(\ell(a ,b)\) is the negative log-likelihood for an observation e.g. for the linear Gaussian case it is \(\frac{1}{2}(a -b)^2\) and for logistic regression it is $ - a \cdot b + \log (1+e^{b})$.  Elastic net is controlled by \(\alpha \in [0,1], \) that bridges the gap between lasso (\(\alpha=1\)) and ridge regression (\(\alpha=0\)). As the true genetic basis of a given trait is generally unknown as well as the LD structure is hard to estimate, we suggest to use a small value for $\alpha $, e.g 0.001. The tuning parameter \(\lambda >0\) controls the overall strength of the penalty and we use 10-fold cross-validation to choose suitable value for $\lambda$. Elastic net approach is implemented in the software 'pyseer' \cite{lees2018pyseer,lees2020improved} focusing on GWAS for bacterial data.

\section{Simulation studies}
\label{sec.simulation}

\begin{figure}
\centering
\includegraphics[height=8cm, width=7cm]{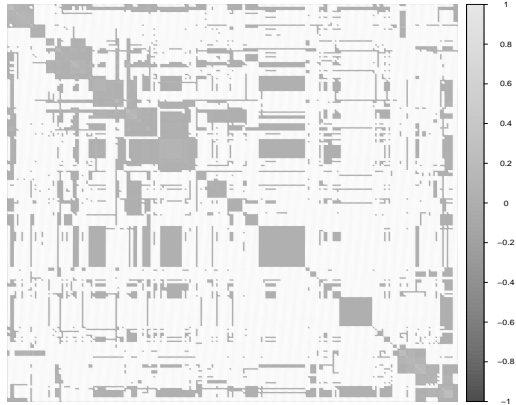}
\vspace*{-1.5cm}
\caption{Sample covariance matrix of the first 100 SNPs covariates in the genotype matrix shows the complex dependence structure present in the \textit{S. pneumoniae} data.}
\label{fg_covarmat}
\end{figure}

We use a real data set of 616 \textit{Streptococcus pneumoniae} genomes collected from Massachusetts, denoted MA data, to create semi-synthetic datasets that incorporate levels of population structure and LD occurring in natural populations (see Figure \ref{fg_covarmat}). The data are publicly available through the article \cite{croucher2015population}. After initial data filtering with standard population genomic procedures (using a minor allele frequency threshold and removing missing data), we obtain a genotype matrix of 603 samples with 89703 SNPs. Using this observed genotype matrix, we simulate the responses/phenotypes through the linear model defined in \eqref{linearmodel}. 

Availability of data and code: The R codes and data used in the numerical experiments are available at:  \url{https://github.com/tienmt/boostingher} .

\subsection{Experimental designs}

We consider the following designs for choosing the causal SNPs (non-zero effect sizes):
\begin{itemize}
\item sparse setting: 100 SNPs are randomly chosen.
\item polygenic setting: 5000 SNPs are randomly chosen.
\item Penicillin resistance -like setting (\cite{lees2016sequence}): 100 SNPs are randomly chosen from 3 genes (pbpX,pbp1A,penA).
\end{itemize}
Given the SNPs, the regression coefficients $\beta^0$ are sampled from the normal distribution $\mathcal{N}(0,1)$. As the true covariance of the genotype matrix is not known, we need to re-normalize the coefficient $\beta^0$ as $\beta = \beta^0  \sqrt{\sigma^2_\varepsilon  h^2/ (\beta^{0\top} \bar{\Sigma} \beta^0 ( 1-h^2 )) } $ to assure that the true corresponding heritability is approximating our target. Here $h^2$ is the target heritability and $\bar{\Sigma} $ is the sample covariance matrix of the genotype matrix and the noise variance is fixed as $ \sigma_{\varepsilon}^2 = 1 $.

The target heritability is fixed as $h^2 = 0.5$. We remind that as true covariance matrix of the genotype matrix is not known, one can only simulate phenotypes from model \eqref{linearmodel} that approximately target the considered heritability. Therefore, we propose to use the ``oracle" estimator, denoted by \textbf{h2aprx}, that is calculated through the formula \eqref{herit:formula-re-write}
\begin{align*}
{\rm h2aprx} = 1 -  \frac{\sigma_{\varepsilon}^2 }{{\rm Var}(y)  } ,
\end{align*}
as a benchmark for comparison. As in simulations the true covariance matrix is not known in our setup, whereas the noise variance is given and thus this estimator provides a solid basis for approximating the true heritability. It is noted that the ``h2aprx" estimator is based on the true simulated values and cannot be used with real data. 

For each setup, we generate 30 simulation runs and report the mean and the standard deviation of heritability estimates for each method across the simulation runs. We compare Elastic net (Enet), HERRA and the boosting versions of HERRA denoted by ``B\_herra" and GCTA method. More specifically, GCTA \cite{yang2011gcta} is a widely used method based on a linear mixed model and maximum (restricted) likelihood estimation. The number of repeated sample splitting is performed with $ B = 50 $ times. The Enet is used with fixed parameter $\alpha = 0.01$ and 10-fold cross validation for choosing the tunning parameter $\lambda $.

\subsection{Results for estimating heritability}
From the results in Table \ref{table_results}, it is clear that the ``oracle" approximates well the target heritability in all designs. Generally, the boosting procedure tends to reduce the variability of the original underlying method it is used in conjunction with, see Table \ref{table_results}, \ref{tb_screening} and \ref{table_results_GCTAmodel}.

Elastic net underestimates the target, which can be explained by the downward bias known to influence  the naive plug-in lasso-type approaches, such as the Elastic net. The effect is due to shrinkage of some of the coefficients corresponding to weak effect towards zero, while such weak effects may still be significant in terms of the total genetic trait variability. However, we would like to note that estimating heritability through Elastic net provides a good lower bound for the heritability, as indicated by the results.

\begin{table}[H]
\centering
\caption{Simulation results with MA data using linear model and the target heritability $h^2 = 0.5$.  The mean and the  standard deviation (in parentheses) of the estimated heritabilities between the simulation replicates are presented.} 
\begin{tabular}{ | p{1.5cm} |  p{2.8cm}| p{2.8cm} |  p{3.8cm} |}
\hline
		& 100 causal SNPs, $ \sigma_{\varepsilon}^2 = 1 $ & 5000 causal SNPs, $ \sigma_{\varepsilon}^2 = 1 $ 
		&  100 causal SNPs from 3 genes, $ \sigma_{\varepsilon}^2 = 1 $
\\ \hline
h2aprx & 0.5004 (.0245) & 0.5085 (.0256)  & 0.4966 (.0227)
		\\
Enet & 0.3585 (.0348) & 0.3770 (.0500) & 0.3546 (.0386)
		\\
HERRA & 0.5619 (.0507) & 0.5583 (.0366)  & 0.5204 (.0483)
		\\
B\_herra & 0.5551 (.0350) & 0.5588 (.0294) &  0.5184 (.0371)
\\
GCTA & 0.3592 (.0309) & 0.3005 (.0270)  & 0.3338 (.0430)
\\ \hline
\end{tabular}
\label{table_results}
\end{table}

On the other hand, HERRA and its boosting version return stable estimates. More specifically, with a proper choice of the screening step (Step 0) as in Table \ref{tb_screening}, HERRA and B\_herra can lead to accurate estimates.This can be anticipated as this approach follows the spirit of the 'oracle' estimator. More specifically, it aims at providing a consistent estimate of the noise variance and thus the corresponding heritability estimate would be also consistent and stable \cite{gorfine2017heritability}. For this reason, the boosting HERRA will be our main focus method in real application in the next section.

In our simulations, GCTA generally did not perform well, most likely due to the sample size being too small for random effects based approaches such as GCTA. We note that,  for unrelated individuals and common SNPs in human studies, GCTA method is recommended with at least 3160 unrelated samples, see \cite{yang2011gcta}. In studies of bacterial phenotypes, it would be uncommon to have access to such large numbers of samples that are at least approximately unrelated.

\subsubsection*{The effect of multiple data splitting}

\begin{figure}[H]
\centering
\includegraphics[height=7cm, width=10cm]{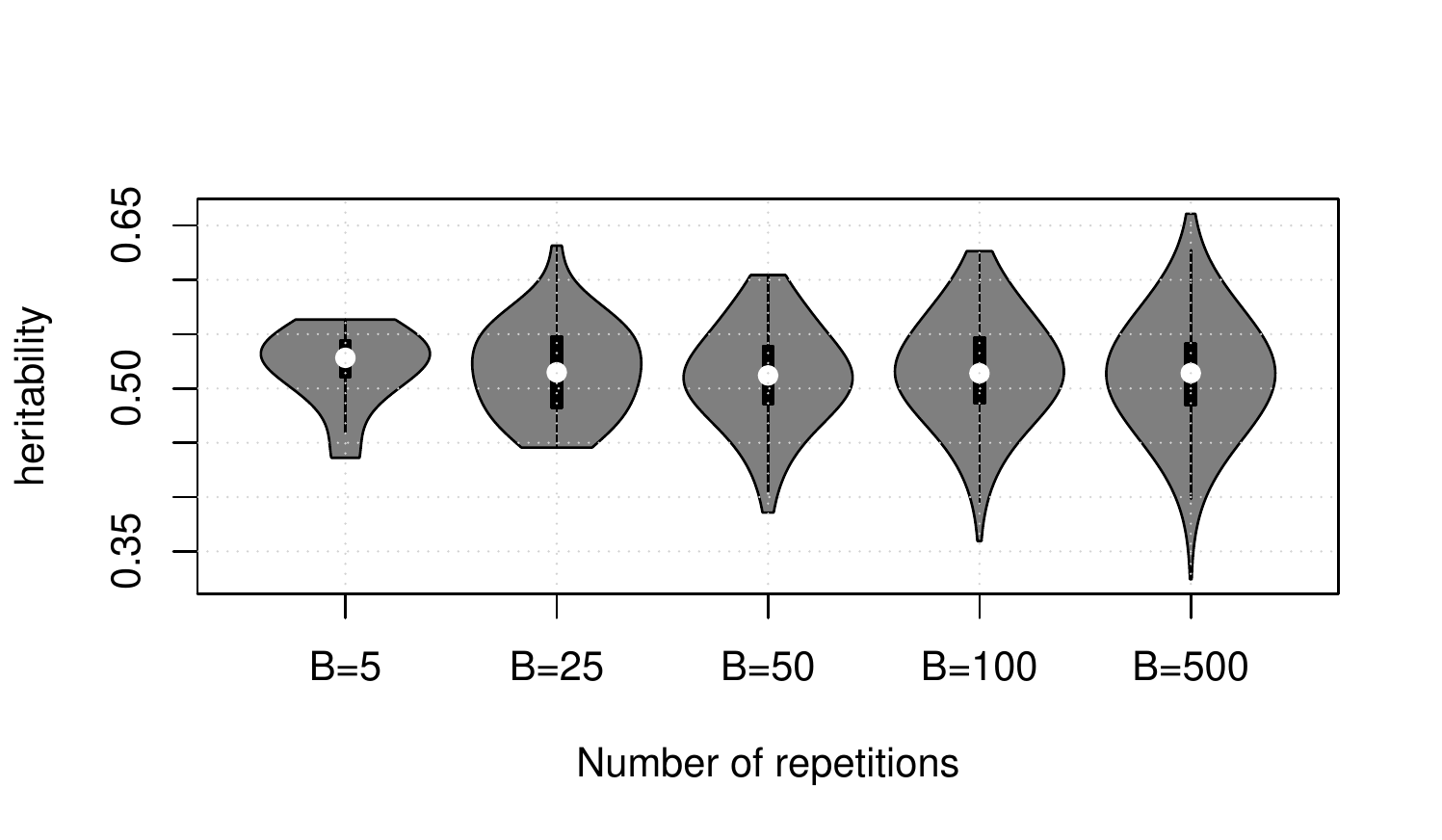}
\vspace*{-.5cm}
\caption{Simulation results with MA data, 100 randomly selected SNPs, $\sigma^2_\epsilon =1 $ and the target heritability $h^2 = 0.5$. Violin plot depicts the distribution of heritability estimates for each chosen $B$, the number of data splittings.}
\label{fg_changingB}
\end{figure}

Clearly, choosing the number of data splittings $B$ is a crucial factor in practice. Here we exemplify that as $B $ increases, the resulting estimated heritabilities concentrate around their mean, see Figure \ref{fg_changingB}. Thus, we suggest to use at least $B \geq 30 $ in practice and $B = 100$ would be a reasonable choice, computational resources permitting. 

\subsubsection*{The effect of the screening step}

We further investigate the effect of reducing the covariates by using the screening step. Different scenarios for 100 randomly selected SNPs with target heritability $h^2 = 0.5$ and $ \sigma_\epsilon^2 = 1 $ are examined, see Table \ref{tb_screening}. More precisely, we further consider 3 scenarios: remove 60\% of the covariates, remove 90\% of the covariates, and only retain top $n+1$ covariates.

It is revealed that using the screening step to reduce the irrelevant covariates not only reduces the dimension of the data, but can also improve the heritability estimation, in particular for the scenario of removing 60\% of the covariates. This fact has also been reported before in the linear mixed model approach in \cite{bonnet2018improving}, where the authors show an improvement of the maximum likelihood estimation. However, if too many covariates are removed, heritability estimation can be inaccurate as in the scenario of keeping only top $n+1$ covariates. 

\begin{table}[H]
\centering
\caption{Simulation results with MA data, 100 randomly selected SNPs, $\sigma^2_\epsilon =1 $ and $h^2 = 0.5$.  The mean and the  standard deviation (in parentheses) of the estimated heritabilities between the simulation replicates are presented. } 
\begin{tabular}{ | p{1.5cm} |  p{4cm}| p{4cm}| p{4cm} |}
\hline
		& remove 60\% covariates & remove 90\% covariates & keep $(n+1)$ covariates
		\\ \hline
Enet & 0.4600 (.0320) & 0.4428 (.0337) & 0.3262 (.0189) 
		\\
HERRA & 0.4921 (.0267) & 0.4788 (.0384) & 0.4063 (.0270) 
		\\
B\_herra & 0.4945 (.0229) & 0.4740 (.0318) & 0.4046 (.0274) 
\\ \hline
\end{tabular}
\label{tb_screening}
\end{table}

\subsubsection*{On the running time}
The running times for default B\_herra on MA data with the splitting step parallelized on 10 CPU cores was 2.335 mins.  More specifically, the screening step took 5.25 secs of the total runtime. In the case of removing 60\% covariates, the running time is significantly reduced to 1.319 mins. The R codes were run on Linux (Redhat 64-bit) with R version 3.6.0 .

\subsection{Simulation results using GCTA model}
We further examine the performances of Enet, HERRA,  B\_herra and GCTA method when the phenotypes are instead simulated from the GCTA model. We remind that GCTA model is a random effect model that is different to the linear model \eqref{linearmodel} and thus we cannot use the 'h2aprx' estimator. The settings for choosing the causal SNPs remain the same as before.

\begin{table}[H]
\centering
\caption{Simulation results with MA data using GCTA model with the true heritability $h^2 = 0.5 $. The mean and the  standard deviation (in parentheses) of the estimated heritabilities between the simulation replicates are presented.} 
\begin{tabular}{ | p{5cm} |  p{2.8cm}| p{2.8cm} | p{3cm} | p{3.8cm} |}
\hline
		& 100  causal SNPs & 5000  causal SNPs
		&  100  causal SNPs from 3 genes
\\ \hline
GCTA &  0.3176 (.0268) & 0.2890 (.0358) & 0.3614 (.0395)
\\
Enet & 0.4014 (.0393) & 0.4018 (.0538) & 0.3965 (.0474)
		\\
HERRA &  0.5248 (.0342) & 0.5142 (.0586) & 0.5246 (.0427)
		\\
B\_herra & 0.5217 (.0260) & 0.5150 (.0447) & 0.5192 (.0339)
\\ \hline
Enet (remove 60\% covariates) & 0.4541 (.0364) & 0.4614 (.0371) & 0.4408 (.0403)
		\\
HERRA (remove 60\% covariates)  & 0.4988 (.0356)  & 0.4941 (.0469) & 0.4966 (.0338)
		\\
B\_herra (remove 60\% covariates)  & 0.5015 (.0260) & 0.4892 (.0426) & 0.4965 (.0306)
\\ \hline
\end{tabular}
\label{table_results_GCTAmodel}
\end{table}

The results, Table \ref{table_results_GCTAmodel}, reveal that HERRA, B\_herra yeild unbiased estimates in GCTA model.  Although underestimate, Elastic net still provides a good lower bound for the true heritability.  Once again, GCTA method underestimates the heritability as the sample is too small.

\section{Heritability of antibiotic resistance in Maela data}
\label{sc_realdata}

To further illustrate the boosting based approach, we apply our procedure to Maela data which represent 3069 \textit{Streptococus pneumoniae} genomes from an infant cohort study conducted in a refugee camp on the Thailand-Myanmar border \cite{chewapreecha2014comprehensive,lees2016sequence}. After some data filtering with standard population genomic procedures (using a minor allele frequency threshold and removing missing data), we obtain a genotype matrix  with 121014 SNPs. We consider resistances to five different antibiotics as the phenotypes: chloramphenicol, erythromycin, tetracycline, penicillin and co-trimoxazole. 

The heritability of the antibiotic resistance phenotype is expected to be high, meaning that the variability stems primarily from the observed genetic differences among these bacteria and that the SNPs available for this particular species/dataset and would include majority of the underlying causal mechanisms for resistance. However, despite that the bacterial isolates are related, it cannot be concluded that the reported estimates refer to \textit{total heritability}, since unmeasured genetic factors are likely to contribute partially to the measured phenotypic variation. We use two different types of resistance phenotypes to investigate their heritability. First we use the binary phenotype corresponding to the labels 'R' or 'S' (stand for 'Sensitive' or 'Resistant') for each bacterial isolate in the cohort. Second, we use a continuous phenotype corresponding to the inhibition zone diameters measured in the lab. These inhibition zone diameters are in practice used to defined whether a sample is sensitive or resistant to an antibiotic. It is however worthwhile noting that the transformation from inhibition zone diameters to labelling a sample 'S' vs 'R' is nonlinear due to the way the inhibition mechanism dynamics in the bacterial culture.

We apply Enet, HERRA, boosting version of HERRA and GCTA method \cite{yang2011gcta} to this data. The results are given in the Table \ref{table:binary_antibiotic} and Table \ref{table:continuous_antibiotic} for the two data types, respectively.

\begin{table}[H]
\centering
\caption{ Heritabilities of antibiotic resistance (binary) phenotypes in Maela data (standard deviation is given in parentheses).} 
\begin{tabular}{ | r |  c| c |c| c|}
 \hline
		& Enet &  HERRA & B\_herra & GCTA
\\ \hline \hline
Chloramphenicol  & 0.4623 & 0.7489 & 0.7617 (.0413) & 0.8257 (.0132)
		\\
Erythromycin  & 0.7979 & 0.9150 & 0.9140 (.0119) & 0.7990 (.0141)
		\\
Tetracycline  & 0.8217 &  0.8899 & 0.8928 (.0113)  & 0.8260 (.0127)
		\\
Penicillin  & 0.7369 & 0.8237 & 0.8280 (.0138)	& 0.6695 (.0228)
		\\
Co-trimoxazole  & 0.5324 & 0.6093 & 0.6340 (.0368) & 0.6005 (.0249)
\\ \hline 
\end{tabular}
\label{table:binary_antibiotic}
\end{table}

\begin{table}[H]
\centering
\caption{ Heritabilities of antibiotic resistance phenotypes using inhibition zone diameters in Maela data (standard deviation is given in parentheses).} 
\begin{tabular}{ | r | c |c |c |c |}
\hline
		& Enet	&  HERRA	& B\_herra		& GCTA
\\ \hline \hline
Chloramphenicol  & 0.5133 & 0.6364 & 0.6337 (.0267) & 0.6837 (.0226)
	\\
Erythromycin  & 0.7350 &  0.8413 & 0.8383 (.0140) & 0.7282  (.0196)
		\\
Tetracycline & 0.7364 &  0.8072 & 0.8435 (.0135) & 0.7514 (.0178)
		\\
Penicillin & 0.8092 & 0.8445 & 0.8462 (.0132) & 0.7123 (.0202)
		\\
Co-trimoxazole  & 0.7104 & 0.7840 & 0.7571 (.0210) & 0.7826 (.0157)
\\ \hline
\end{tabular}
\label{table:continuous_antibiotic}
\end{table}

As a broad summary, heritabilities of these five antibiotic resistances are high, as expected, whether using binary or continuous phenotypes. However, we would like to note that the results for binary responses are on the observed scale (0/1 resistance status),  as we are not able to transform them into the underlying threshold model, see \cite{gorfine2017heritability}. The Elastic net method yields an important insight by providing a lower bound on the heritability of these antibiotic resistances. For continuous phenotypes, it is at least 51\% for chloramphenicol, at least 73\% for erythromycin, at least 73\% for tetracycline, at least 80\% for penicillin and at least 71\% for co-trimoxazole. 

Interestingly, B\_herra yields consistent results with GCTA method. However, the result for heritability of penicillin by GCTA is lower than the one from Enet method while boosting HERRA is not, see Table \ref{table:continuous_antibiotic}.

\section{Discussion and conclusions}

In this paper, we provide a general framework 'boosting heritability' for making inference about heritability. The main ingredient of 'boosting heritability' is a multiple sample splitting strategy. This strategy allows one to employ a variable selection step to remove irrelevant covariates that do not contribute to the variability of a trait and thus produce a reliable estimate of heritability. Moreover, by repeating sample splitting many times, this strategy makes sure that different latent structures are taken into account in both selection and estimation steps. 

Numerical comparisons of different methods together with our proposal for estimating heritability in linear (fixed-effect) model draw a systematic picture on the behaviour of the current approaches when focusing on an application to bacterial GWAS. The results on real data suggest that the observed variability of the five studied antibiotic resistances is mainly due to the variability in the observed genetic factors, while some unexplained variation still remains.

Succeeding in improving and stabilizing HERRA \cite{gorfine2017heritability}, ``boosting heritability" framework still preserves its advantages that are able to deal with the dichotomous, time-to-event or age-at-onset traits. Moreover, boosting heritability procedure is also applicable for random-effect model where the heritability estimation step (Step 3 in Algorithm \ref{Boostingheritability}) is done by using a random effect method as in \cite{li2019reliable}. These would be possible new research directions for the future.

Furthermore, our boosting heritability procedure uses a simple aggregation to combine the estimates that is to use their arithmetic mean. Other types of aggregation, see e.g \cite{renaux2020hierarchical,buzdugan2016assessing}, could also be used and further examined in future works.

\section*{Acknowledgments}
The authors would like to thank the editor and two anonymous referees who kindly reviewed the earlier version of this manuscript and provided valuable suggestions and enlightening comments. T.T.M and J.C. would like to thank John A Lees for his useful discussion on GWAS and heritability.

\section*{Funding}
This research was supported by the European Research Council grant no. 742158.

\section*{Availability of data and materials}
The R codes and data used in the numerical experiments are available at:  \url{https://github.com/tienmt/boostingher} .

\section*{Ethics approval and consent to participate}
Not applicable.

\section*{Competing interests}
The authors declare that they have no competing interests

\section*{Consent for publication}
Not applicable.

\section*{Authors' contributions}
Conceptualization: T.T.M. and J.C. . Formal analysis: T.T.M.. Data curation: P.T. . Methodology, T.T.M.. Writing: original draft, T.T.M.; review and editing, all authors. Funding acquisition: J.C..

\bibliographystyle{apalike}

\appendix
\subsection*{Additional simulations}
\begin{table}[H]
\centering
\caption{Simulation results with MA data using linear model and the target heritability $h^2 = 0.5$ (standard deviation is given in parentheses).} 
\begin{tabular}{ | p{1.5cm} |  p{2.8cm}| p{2.8cm} | p{2.8cm} | }
\hline
		&  100 causal SNPs, $ \sigma_{\varepsilon}^2 = 10^2 $
\\ \hline
h2aprx &  0.5026 (.0226) 
		\\
Enet & 0.3724 (.0455) 
		\\
HERRA & 0.5527 (.0494) 
		\\
B\_herra & 0.5467 (.0378) 
\\
GCTA & 0.3272 (.0386)
\\ \hline
\end{tabular}
\end{table}

\subsection*{On details on running time}
\begin{verbatim}
> sessionInfo()
R version 3.6.0 (2019-04-26)
Platform: x86_64-redhat-linux-gnu (64-bit)
Running under: Red Hat Enterprise Linux

Matrix products: default
BLAS/LAPACK: /usr/lib64/R/lib/libRblas.so

locale:
 [1] LC_CTYPE=C                 LC_NUMERIC=C              
 [3] LC_TIME=en_US.UTF-8        LC_COLLATE=en_US.UTF-8    
 [5] LC_MONETARY=en_US.UTF-8    LC_MESSAGES=en_US.UTF-8   
 [7] LC_PAPER=en_US.UTF-8       LC_NAME=C                 
 [9] LC_ADDRESS=C               LC_TELEPHONE=C            
[11] LC_MEASUREMENT=en_US.UTF-8 LC_IDENTIFICATION=C       

attached base packages:
[1] parallel  stats     graphics  grDevices utils     datasets  methods  
[8] base     

other attached packages:
[1] doMC_1.3.6       iterators_1.0.12 foreach_1.4.7    glmnet_3.0-2    
[5] Matrix_1.2-17   

loaded via a namespace (and not attached):
[1] compiler_3.6.0   codetools_0.2-16 grid_3.6.0       shape_1.4.4     
[5] lattice_0.20-38 

\end{verbatim}

\end{document}